# Half-lives of $^{101}$Rh and $^{108}$Ag$^m$

Howard A. Shugart[1,2a], Edgardo Browne[2], and Eric B. Norman*[2,3]

**ABSTRACT:** Using long-term gamma-ray spectroscopy with high resolution germanium detectors and a ratio method, the half-lives of $^{101}$Rh and $^{108}$Ag$^m$ have been determined to be 4.07 $\pm$ 0.05 years and 448 $\pm$ 27 years, respectively. These results are compared with previously reported values.

* Corresponding author
[a] Deceased
[1] Physics Department, University of California, Berkeley, CA 94720 U.S.A.
[2] Nuclear Science Division, Lawrence Berkeley National Laboratory, Berkeley, CA 94720 U.S.A.
[3] Nuclear Engineering Department, University of California, Berkeley, CA 94720 U.S.A.

**Introduction**

One of the defining characteristics of each radioactive nucleus is its half life. Observed nuclear half lives range from tiny fractions of a second to much greater than the age of the universe. The ease with which such half lives can be measured depends on their values. Half lives in the range from microseconds to months can easily be determined by direct counting methods. A simple example of this method is shown in Figure 1, where the half life of $^{137}Ba^m$ was measured by counting the number of 662-keV gamma rays observed as a function of time. On the other hand, half lives in the range from thousands to billions of years can also be fairly easily determined by measuring the specific activity of a sample containing the nucleus of interest. Examples of the use of this method include the determinations of the half life of $^{176}Lu$ by Norman (1980) of $(4.08 \pm 0.24 \times 10^{10}$ y) and that of $^{138}La$ by Norman and Nelson (1983) of $(1.06 \pm 0.03 \times 10^{11}$ y). It is nuclei with half lives in the range of a few years to several thousand years that are actually the most difficult to determine accurately. Two such nuclei are $^{101}Rh$ and $^{108}Ag^m$ whose half lives have been studied a number of times in the past. However, previous half-life measurements of $^{101}Rh$ and $^{108}Ag^m$ have yielded discrepant and inconsistent results, which are summarized in Table 1.

Table 1. Reported half lives for $^{101}Rh$ and $^{108}Ag^m$ (in years).

**$^{101}$Rh**

| | |
|---|---|
| $5 \pm 1$ | Farmer (1955) |
| $10 \pm 3$ | Perrin (1960) |
| $3.0 \pm 0.4$ | Hisatake et al. (1965) |
| $3.3 \pm 0.3$ | Evans and Naumann (1965) |

**$^{108}$Ag$^m$**

| | |
|---|---|
| $310 \pm 132$ | Vonach et al. (1969) |
| $127 \pm 21$ | Harbottle (1970) |
| $418 \pm 15$ | Schotzig et al. (1992) |
| $437.7 \pm 8.8$ | Schrader (2004) |

As a result of these inconsistent results, we decided to perform long term gamma-ray counting measurements of these two isotopes in order to better determine their respective half lives. Due to the lengthy time period required for these measurements, we decided to use a ratio method (described below) in order to reduce possible systematic effects due to variations in detector and/or electronics performance over time.

**Experimental Method**

To produce $^{101}$Rh, a metallic rhodium foil 1.27x1.27x0.127-cm was bombarded with 40 MeV protons from Lawrence Berkeley National Laboratory's 88" cyclotron. The integrated charge delivered to the water-cooled target was 3.94 x $10^{-3}$ coulombs in 143 minutes resulting in an average current of about 0.46 microamperes. The proton beam exited the Rh foil with about 24 MeV energy in order to emphasize the (p,3n) (p,4n) and the (p,pn) or (p,d) reactions. Following the bombardment, the target was counted using a shielded "25% efficient" high-purity germanium detector. Data were collected using an ORTEC PC-based data acquisitions system. Over a 6 year period, approximately 250 three-day long gamma-ray spectra were taken approximately once per week at a distance from the detector of 47 cm. Although gamma rays from the decays of $^{100}$Pd (3.63 d) $^{100}$Rh (20.8 h) and $^{101}$Rh$^m$ (4.26 d) were initially observed, after 66 days these had decayed, and only the long lived isotopes $^{101}$Rh, $^{102}$Rh$^g$, and $^{102}$Rh$^m$ were detected. A typical spectrum observed in three days of counting this target is shown in Figure 2.

For our second half life measurement, we obtained a source of $^{108}$Ag$^m$ from Richard Helmer of Idaho National Laboratory and mounted it together with sources of $^{133}$Ba and $^{44}$Ti near a shielded 3.6-cm diameter and 1.3-cm thick planar germanium detector. Data were collected automatically using a separate ORTEC PC-based data acquisitions system. Gamma-ray spectra were collected in one-week long intervals, written to disk, cleared, and then started again. A typical gamma-ray spectrum obtained from one week of counting is shown in Figure 3. This experiment ran for approximately three years.

**Analysis and Results**

Rather than attempting to perform absolute measurements of the half lives of $^{101}$Rh and $^{108}$Ag$^m$, in both cases we used reference isotopes that appear in our spectra to determine relative half lives. In the case of $^{101}$Rh, we made use of $^{102}$Rh$^m$ whose half life was taken to be 3.742 $\pm$ 0.010 y as reported by Shibata et al. (1998). Net peak areas of $^{101}$Rh gamma rays at 127, 198, and 325 keV were carefully extracted from our spectra and compared to those of the 696-keV peak produced by the decay of $^{102}$Rh$^m$. The ratios of $^{101}$Rh/$^{102}$Rh$^m$ gamma-ray peak areas as a function of time are shown in Figure 4. By using such ratios, possible variations in detector and/or electronics performance over the long counting period are largely cancelled out. The fact the slopes of all of these plots are positive implies that the half life of $^{101}$Rh is longer than that of $^{102}$Rh$^m$. Least squared fits were performed to each of these data sets to obtain an effective "growth rate". From the difference between each of these growth rate values and the well established decay rate of $^{102}$Rh$^m$, the decay rate of $^{101}$Rh was determined. Our combined result for the half life of $^{101}$Rh is 4.07 $\pm$ 0.05y which agrees with that of Farmer (1955) but disagrees with all of the other reported values.

In a similar way, for the determination of the half life of $^{108}$Ag$^m$, we made use of $^{133}$Ba as our reference isotope. Its half life has been well determined to be 10.551 $\pm$ 0.011 y (Khazov et al., 2011). We carefully extracted the net peak areas of the $^{133}$Ba 356-keV gamma ray and those of the $^{108}$Ag$^m$ gamma at 434 keV. The fact that the energies of these two gamma rays are nearly equal means that both energy and time dependent detector and electronics variations largely cancel out in the ratio of the two counting rates. The ratio of these net peak areas as a function of time is shown in Figure 5. A least squares fit was made to this data and the effective decay rate of this ratio was determined to be (6.4146 $\pm$ 0.0063 )x$10^{-2}$/year. The difference between this effective decay rate and that of $^{133}$Ba was then used to determine the $^{108}$Ag$^m$ decay rate. Using this analysis method, the half life of $^{108}$Ag$^m$ was extracted from our data. After one year of counting, a preliminary result for the half life of $^{108}$Ag$^m$ was reported by Norman (2005) to be 440 $\pm$ 84 y. Here we present our final result for the half life of $^{108}$Ag$^m$ from a total of three years of counting to be 448 $\pm$ 27 y. As a check on our technique, we performed a similar ratio analysis between the 356-keV $^{133}$Ba gamma ray and that of the 1157-keV gamma ray produced by the decay of $^{44}$Ti. Our resulting half life for $^{44}$Ti is 58 $\pm$ 1 y, in good agreement with the accepted value of 59.1 $\pm$ 0.3 y (Chen et al., 2011). Our result for the half life of $^{108}$Ag$^m$ is in good agreement, although less precise than those of Schotzig et al. (1992) and Schrader (2004) whose measurements were carried out over 20 years.

**Acknowledgements**

The authors wish to thank the 88" cyclotron crew for their conscientious and helpful attention to all details of the rhodium bombardment. This material is based upon work supported in part by the U. S. Department of Energy National Nuclear Security Administration under Award Number(s) DENA0000979 and by the Director, Office of Energy Research, Office of High Energy and Nuclear Physics, Division of Nuclear Physics, of the U.S. Department of Energy under Contract No. DE-AC02-05CH11231.

**Figure Captions**

1. Half life of $^{137}Ba^m$ determined by counting the number of 662-keV gamma rays observed from a source as a function of time.

2. Typical gamma-ray spectrum observed during a 3-day counting period from our $^{101,102}Rh$ sample.

3. Typical spectrum observed in one week of counting our $^{133}Ba/^{44}Ti/^{108}Ag^m$ sources.

4. Ratios of observed $^{101}Rh$ gamma-ray count rates to that of the 697-keV gamma ray from the decay of $^{102}Rh^m$ as a function of time.

5. Ratio of the observed count rate of the 356-keV $^{133}Ba$ gamma ray to that of the 434-keV $^{108}Ag^m$ gamma ray as a function of time.

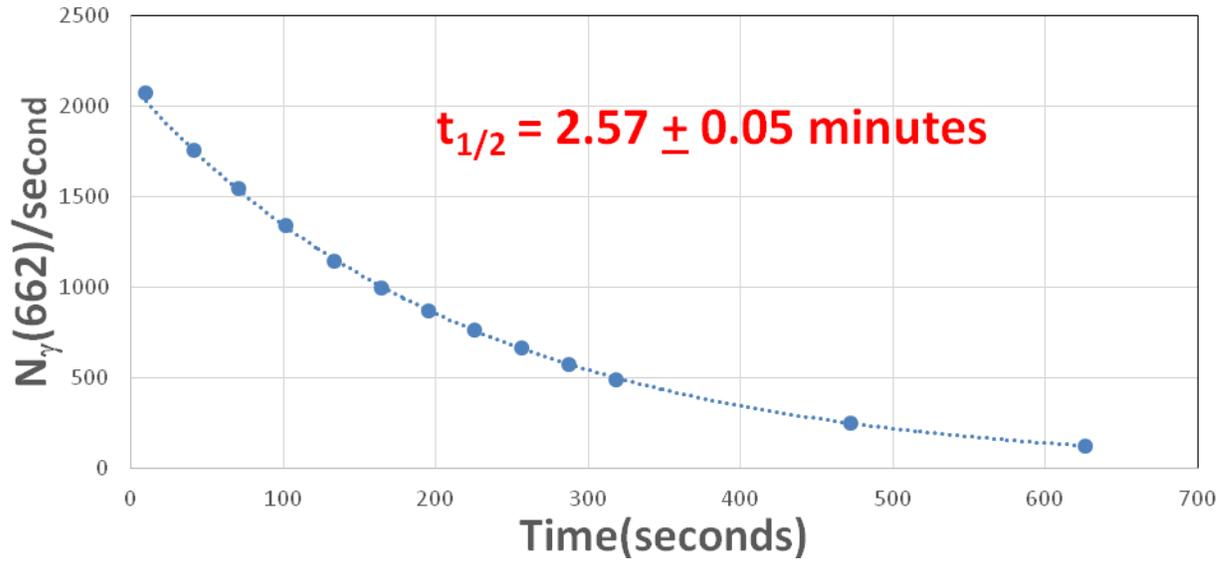

Figure 1

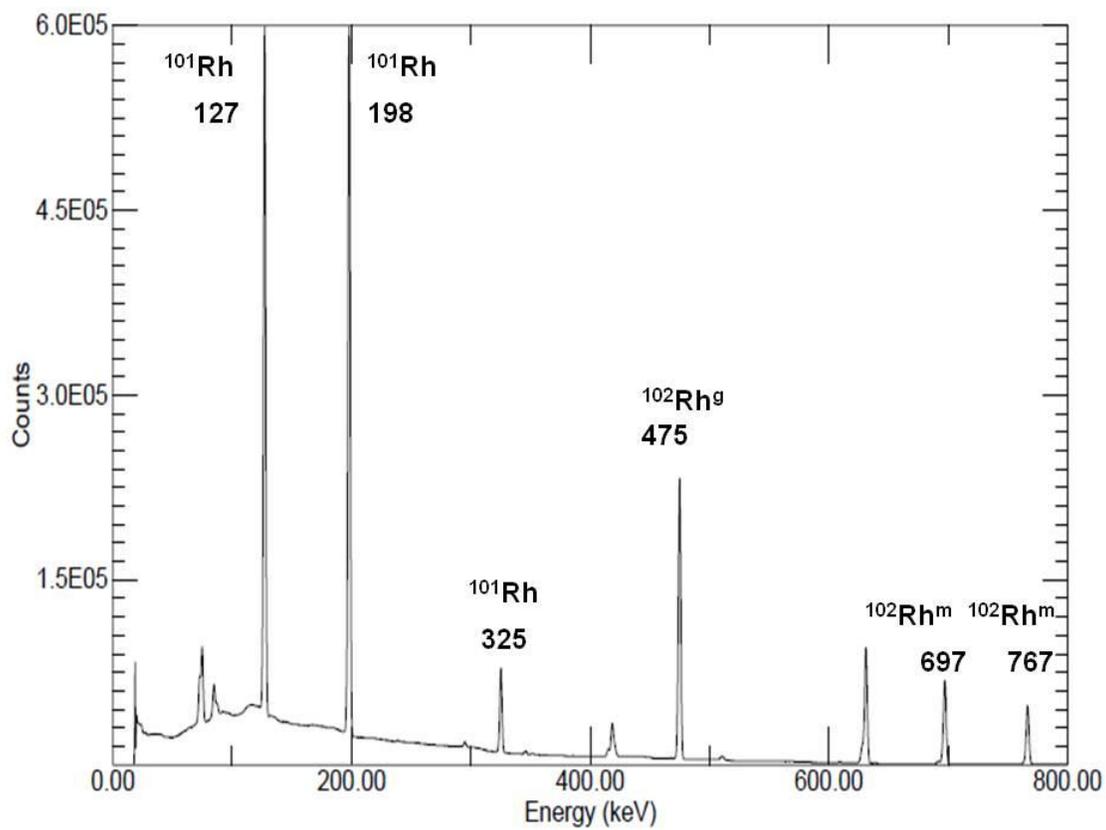

Figure 2

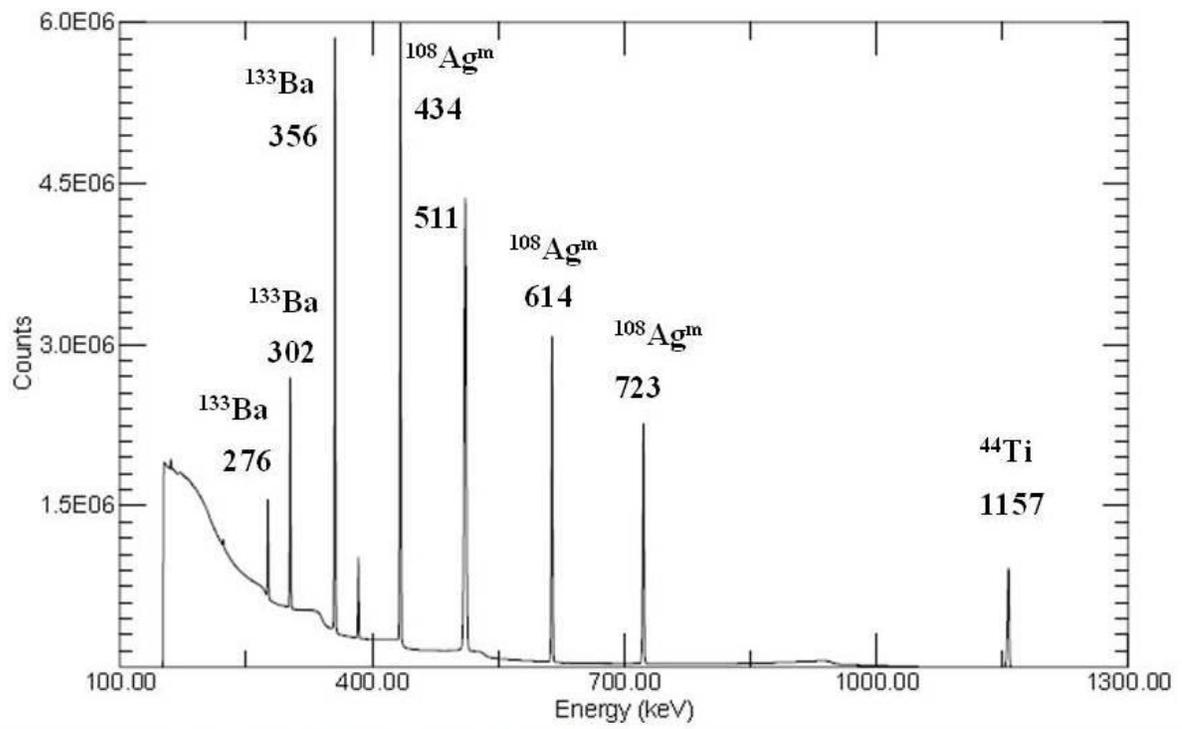

Figure 3

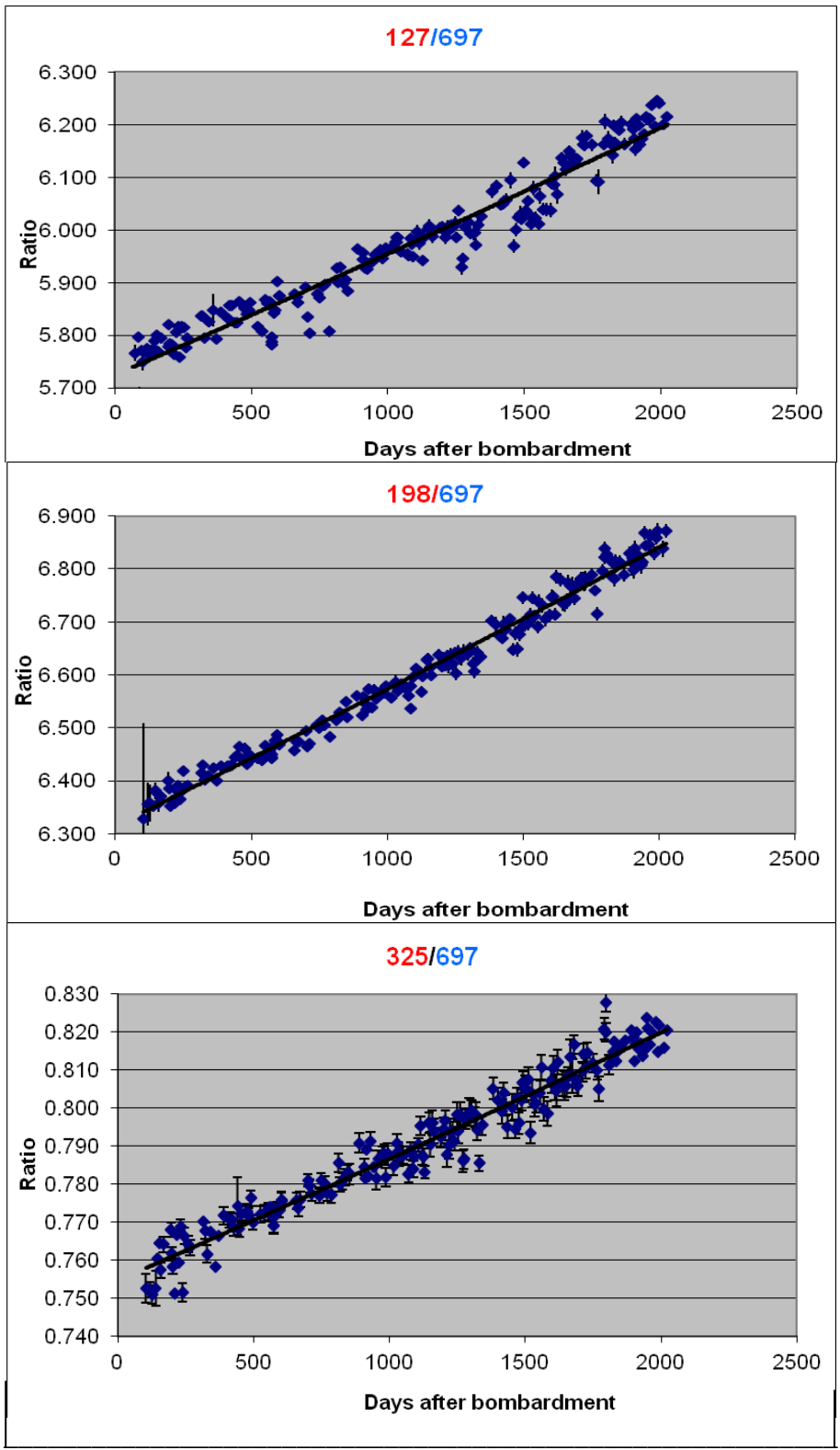

Figure 4

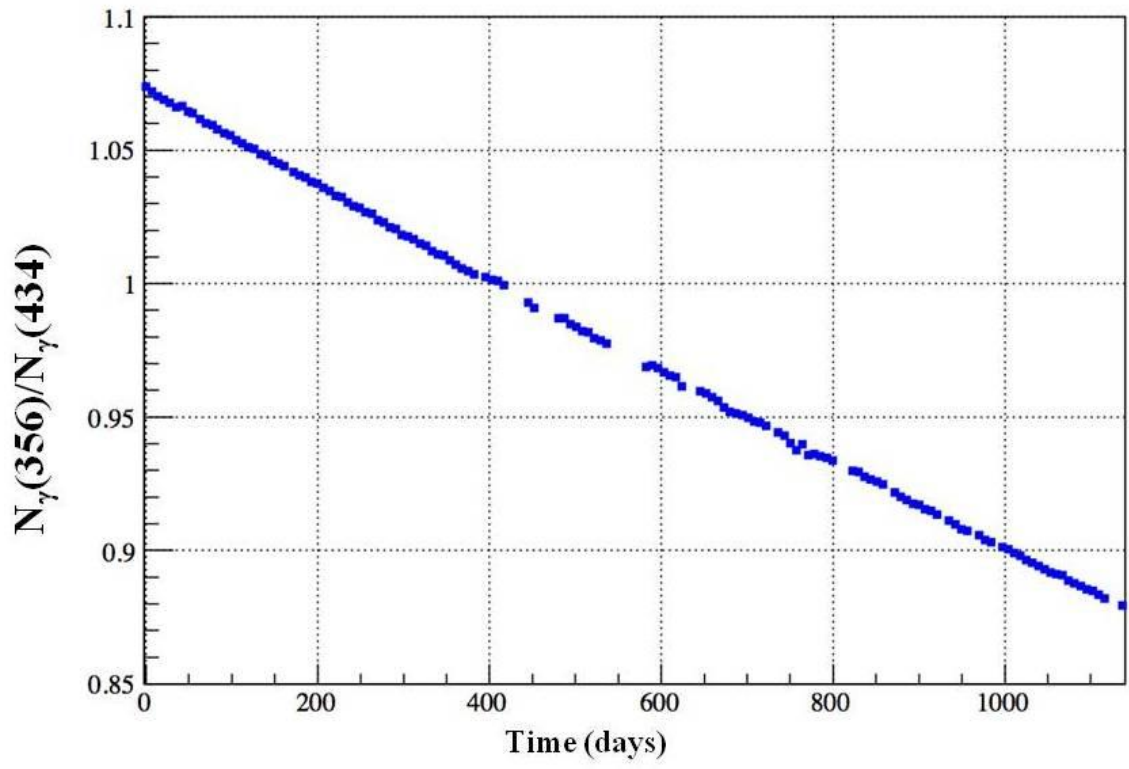

Figure 5